\documentstyle[twocolumn,prl,epsf,aps]{revtex}
\begin{document}
\draft
\title{Synchronisation of time--delay systems}
\author{Martin J. B\"unner\thanks{published in Phys. Rev. E {\bf 58} (1998)
R4072,  e--mail: buenner@mpipks-dresden.mpg.de}
and Wolfram Just\thanks{e--mail: wolfram@arnold.fkp.physik.th-darmstadt.de}}
\address{Max--Planck Institute for Physics of Complex Systems,
N\"othnitzer Stra\ss e 38, D--01187 Dresden, Germany}
\date{February 26, 1998}
\maketitle
\begin{abstract}
We present the linear-stability analysis of synchronised states
in coupled time--delay systems. There exists a synchronisation threshold,
for which we derive upper bounds, 
which does not depend on the delay time. We prove that at least for
scalar time--delay systems synchronisation is achieved
by transmitting a single scalar signal, even if the synchronised solution
is given by a high--dimensional chaotic state with a large number of positive
Lyapunov--exponents. The analytical results are compared with
numerical simulations of two coupled Mackey--Glass equations.
\end{abstract}
\newcommand{\be}{\begin{equation}\label}
\newcommand{\ee}{\end{equation}}
\newcommand{\bea}{\begin{eqnarray}\label}
\newcommand{\eea}{\end{eqnarray}}
The problem of synchronisation of dynamical systems is one of the classical
fields in engineering science \cite{synchronisation}. Recently, renewed 
interest in this field was
stimulated in connection with the synchronisation of chaotic motion.
Especially, the potential applicability for communication 
has attracted much research in recent years \cite{communication}. 
Yet, there are a lot of
results available concerning
the synchronisation of low--dimensional chaotic systems, theoretical as well
as experimental  \cite{low-dim}. 
Contrary, the synchronisation of high--dimensional chaotic systems with 
possibly a large number of positive Lyapunov--exponents remains open.
From the point of view of numerical simulations
the synchronisation of specific
high--dimensional chaotic systems has been achieved
\cite{high-dim}, while,  to our best knowledge, rigorous results,
e.~g.~concerning sufficient synchronisation conditions, are still lacking. 
While it has been proved recently by Stojanovski et al. \cite{Stojanovski97}
that the synchronisation of high-dimensional chaotic states can be in
principle achieved with a single transmitted variable, the problem
of finding the appropriate coupling of the two systems remains open.
For that reason we address in this paper the question of the
synchronisation of coupled identical time--delay systems.
We focus on time--delay systems, since on the one hand 
it is well established that these systems are prominent examples of
high--dimensional chaotic motion with a large number of positive
Lyapunov--exponents \cite{time-delay}, and one the other hand 
synchronisation of Mackey--Glass type electronic oscillators 
has been reported from the experimental point of view \cite{Tamasevicius}.

Let us consider a fairly general theoretical model and
investigate the stability problem of a synchronised state.
For that purpose consider two identical 
arbitrary scalar time--delay systems with a symmetric coupling 
\bea{2-a}
\dot{x}&=&F(x,x_{\tau})-K(x-y) \quad , \nonumber\\
\dot{y}&=&F(y,y_{\tau})-K(y-x) \quad ,
\eea
where we adopt the notation $x_{\tau}:=x(t-\tau)$ to indicate the
time--delayed variables. We specialise from the beginning to the frequently 
analysed case that
the coupling is bi--directional and acts additive to the single dynamical 
system. However,
we stress that the subsequent considerations apply with minor modifications
to much more general situations, e.~g.~to vector--type variables, to systems 
with much more general delay terms, or to a non--additive coupling, 
as long as the coupling vanishes in the
synchronised state $x(t)\equiv y(t)$. But we think, that the
choice made in eq.(\ref{2-a}) makes our arguments more transparent.

Let $z$ denote the synchronised solution, i.e. 
$\dot{z}=F(z,z_{\tau})$. Considering deviations from that state according
to $x=z+\delta x$, $y=z+\delta y$ and performing a linear stability analysis,
we obtain for the deviation $\Delta:= \delta y -\delta x$ from the 
synchronised state the linear differential--difference equation
\be{2-b}
\dot{\Delta} = \alpha(t) \Delta + \beta(t) \Delta_{\tau} \quad .
\ee
Here, the time--dependent coefficients are given in terms of the
synchronised solution as $\alpha(t)=\partial_1 F(z,z_{\tau})-2K$ and
$\beta(t)=\partial_2 F(z,z_{\tau})$, where the symbol
$\partial_{1/2}$ denotes the derivative with respect to the 
first/second argument. A superficial inspection of eq.(\ref{2-b})
might suggest that the synchronised solution is stable if
$\alpha(t)$ is ''sufficiently negative''. In fact, we will make this 
statement rigorous in what follows.
Suppose the coefficients are bounded in the sense that 
$\alpha(t)\leq -a<0$ and $|\beta(t)|\leq b$ holds for some
fixed values $a$ and $b$. 
Since the equation is linear, it is sufficient to analyse the 
solution with the special initial condition
$\Delta(0)=1$, $\Delta(t) \equiv 0, t<0$.
The general case follows by a simple integration. 
There are different
ways to estimate the stability of the trivial solution, $\Delta(t) \equiv 0$, 
of eq.(\ref{2-b}). 
Here we use the fact that for scalar quantities a simple 
closed analytical formula for the solution can be written 
down. One just integrates the linear equation (\ref{2-b}) in the
time intervals $[N\tau,(N+1)\tau]$ and considers the delay term as
an inhomogeneous part. By this continuation method (cf.\cite[p.45]{bell})
the full solution is obtained as
\bea{2-c}
\Delta(t)&=& e^{\int_0^t \alpha(t')\, dt'} + 
\int_{\tau}^t dt_1 \beta(t_1) e^{\int_{I_1} \alpha(\theta)\, d\theta}
\nonumber\\
&+& \int_{2\tau}^t dt_1 \int_{2\tau}^{t_1} dt_2
\beta(t_1) \beta(t_2-\tau) 
e^{\int_{I_2} \alpha(\theta)\, d\theta} + \ldots\nonumber\\
&+& \int_{N \tau}^t dt_1 \int_{N\tau}^{t_{1}} dt_2 
\ldots \int_{N\tau}^{t_{N-1}} dt_N \nonumber \\
& & \beta(t_1) \beta(t_2-\tau) \ldots \beta(t_N-(N-1)\tau)
e^{\int_{I_N} \alpha(\theta)\, d\theta} \nonumber\\
& & \mbox{ for } N\tau \leq t \leq (N+1) \tau
\quad .
\eea
Here, the domains of integration for the exponents are given by
$I_k:=[0,t]/([t_1,t_1-\tau]\cup[t_2-\tau,t_2-2\tau]\cup \ldots \cup
[t_k-(k-1)\tau,t_k-k\tau])$. An upper bound for 
$|\Delta(t)|$ is obtained, if the maximal values $\alpha(t)=-a$ and 
$\beta(t)=b$ are inserted into eq.(\ref{2-c}). But then, the expression
reduces to a solution $\Gamma$ of the differential--difference equation 
with constant coefficients
\be{2-d}
\dot{\Gamma} = -a \Gamma+ b\Gamma_{\tau} \quad.
\ee 
Hence, a solution of eq.(\ref{2-d}) yields an upper bond for $|\Delta(t)|$. 
But the last equation is easily solved by a Laplace
transformation (cf.\cite{bell}) or loosely speaking by an exponential
ansatz $\Gamma(t)= e^{s t}$. Since the corresponding eigenvalues obey
$s=-a+ b \exp(-s \tau)$, negative real parts, i.~e.~stability,
occur if and only if $a>b$. This inequality yields an upper bond $K_+$
for the critical coupling strength beyond which synchronisation is
achieved. If we take the definitions of $\alpha(t)$ and $\beta(t)$ into 
account it reads explicitely
\be{2-da}
K_+=1/2 [ \max_t \partial_1 F(z,z_{\tau}) + 
\max_t|\partial_2 F(z,z_{\tau})|] \quad .
\ee
We note as a by--product that eq.(\ref{2-d})
may be viewed as a kind of Grownwall--like lemma \cite{Grownwall} for the 
time--dependent equation (\ref{2-b}).

In what follows, we compare our analytical result to numerical simulations. 
We specialise to the Mackey--Glass system, i.~e.
\be{2-e}
F(x,x_{\tau})=-x + \frac{a x_{\tau}}{1+x_{\tau}^{10}} \quad .
\ee
In order to investigate the properties of the synchronisation mechanism 
by numerical methods, we chose the distance between trajectories 
as a suitable measure. For that reason the quantity
\begin{equation}
\label{abstand} 
D_T(t)=\int_{t}^{t+T} \mid x(t')-y(t') \mid dt', 
\end{equation}
which of course depends on the
range of averaging $T$ and the point of reference $t$ was analysed.

We used a Runge--Kutta algorithm of fourth order with step size $0.1$. 
The simulation, which have been performed for the parameter value $a=3$,
started with $K=0.35$. A constant initial condition for $x$ and $y$, which
differs by an amount of $10^{-3}$ has been chosen.
The system was allowed to relax for a time $t=80\tau$. After that, the 
distance $D$ was integrated on a trajectory of the length $T=80\tau$. 
For the next value of coupling strength $K$ we again distorted the last
state of the trajectory  by adding an amount of $10^{-3}$ to the 
$y$--coordinates and used it as initial condition.
We performed the computation for increasing as well as
decreasing coupling constant. Fig.\ref{fig1} summarises our findings.
\begin{figure} 
\caption[]{Distance $D$ for two coupled Mackey--Glass systems for $\tau=10$
(dashed line) and $\tau=100$ (solid line). \label{fig1}}
\end{figure}
\begin{figure} 
\caption[]{Time series for $\tau=10$ and different values of the coupling
(a) $K=0.2595$, (b) $K=0.250$, (c) $K=0.240$. \label{fig2}}
\end{figure}

For $\tau=10$ we observe distinct jumps,
indicating that the system switches between coexisting periodic states with a
pronounced hysteresis. For $\tau=100$, the overall behaviour of the system
appears to be quite similar as for $\tau=10$, except that no switching and no
hysteresis is observed. Within the resolution of the graphics the same
behaviour has already been observed for a smaller value $\tau=50$.
From the numerical simulations the synchronisation threshold is estimated as
$K_c(\tau=10) \approx 0.24$, $K_c(\tau=100) \approx 0.28$.
If we evaluate our analytical estimate eqs.(\ref{2-da}) and (\ref{2-e})
using upper bounds for the derivatives we obtain values which differ
by an order of magnitude but are independent of the delay time
$\tau$, $K_+=(81 a/40 -1)/2=2.5375$. Since we have applied a rather
graceful rigorous estimate such a discrepancy is far from being
astonishing.

In order to understand the dynamics in the vicinity of
the synchronisation threshold, time traces of the difference $x(t)-y(t)$ 
have been computed (cf.~fig.\ref{fig2}).
Slightly below the synchronisation threshold $K_c$ we observe an intermittent
behaviour very similar to on--off intermittency \cite{intermittency}.
Additionally, we investigated the distribution of laminar phases and turbulent
phases under variation of the coupling strength $K$ and the delay time $\tau$,
which we present in fig. \ref{fig3}. To this end, the
distance $D_{\tau}(t)$ of the two systems in the phase space has been computed 
on an intervall of length $\tau 10^6$. Then, the length of the laminar phases ($D_{\tau} \le 0.10$) 
and turbulent phases ($D_{\tau} > 0.10$) were recorded. We observe a power-law scaling of the 
distribution $P_l$ of the laminar phases over
a wide range, $P_l \propto t^{-\alpha_l(\tau)}$, where the exponent $\alpha_l$ depends slightly
on the delay time. In the low-dimensional chaotic case, for $\tau=10.0$, we observe
$\alpha(10)=1.50$ in agreement with the value predicted by the scenario of 
on-off intermittency. In the high-dimensional chaotic cases, for increasing delay time, 
we observe a decreasing exponent: $\alpha(30.0)=1.41, \alpha(50.0)=1.38, \alpha(100.0)=1.27$,
indication that there might be deviations from on--off intermittency. The distribution
of the turbulent phases follow a $P_t \propto t^{-\alpha_t(\tau)}$-scaling for high enough $\tau$.
\begin{figure} 
\caption[]{(a) Distribution of laminar phases, (b) distribution of turbulent phases 
for  $\tau=100$. (solid line: $K=K_c=0.28$; circles: $K=0.26$, dotted line: $K=0.24$, 
dashed line: shifted power-law fit for $K=K_c=0.28$ with 
$\alpha_{l}= 1.27$, and $\alpha_{t}= 2.19$
\label{fig3}}
\end{figure}
Although these findings are at the first sight not surprising, a detailed 
analysis is required, with special focus on the high dimensionality of
the dynamics. Details will be reported elsewhere. 

We conclude with a remark, how our results depend on noise or other 
imperfections which are present in realistic systems. In fact, in order to
apply a concept like synchronisation such perturbations have to be small and
we may assume a general linear dependence. Formally
such contributions are introduced into eq.(\ref{2-a}) by adding the two
terms $G(x,x_{\tau}) \xi$ and $G(y,y_{\tau}) \eta$, where
$\xi$ and $\eta$ denote for example realisations of a noise. 
Considering the perturbations of the same order of magnitude like 
the deviation from the unperturbed synchronised state and proceeding as above
we finally end up with
\be{2-f}
\dot{\Delta} = \alpha(t) \Delta + \beta(t) \Delta_{\tau}
+G(z,z_{\tau}) (\eta-\xi) \quad ,
\ee
which differs from eq.(\ref{2-b}) just by an inhomogeneous contribution.
The theory of linear difference--differential equations tells us \cite{bell}
that eq.(\ref{2-f}) inherits its stability properties from the
corresponding homogeneous system (\ref{2-b}) except that the 
perturbations cause fluctuations around the unperturbed synchronised state.
Whenever the perturbations are so large, that contributions beyond the linear
order have to be taken into account, one has to resort to different methods.
One of these cases, which are also relevant from the experimental point of 
view, is given by the synchronisation of nearly-identical 
time-delay systems. Since, in this case, no strict synchronised solution
$x \equiv y$ exists, one has to rely on more general concepts, such as the 
generalised synchronisation \cite{generalized}. 

In summary, we emphasise that an analytical upper bond for the solution of 
eq.(\ref{2-b}) is obtained if one replaces the time dependency of the 
coefficients by their extreme values. One might get better estimates 
in special cases. In particular one might argue, that eq.(\ref{2-d}) 
already determines the stability, if the time averages of the 
coefficients are inserted. This statement is in fact true if either 
the coefficients are 
periodic functions of time with the delay $\tau$ being an integer multiple of 
the period or, if the coefficients are almost constant 
(cf.\cite[p.277]{bell}). Whether the general case can be treated 
by this refined estimate remains open. Nevertheless we have shown, that
for sufficiently large coupling constant $K$ the synchronised solution
of eq.(\ref{2-a}) becomes stable, whenever $|\delta_{1/2} F(z,z_{\tau})|$
are uniformly bounded. In particular the critical coupling strength
remains bounded even in the limit of large delay times. i.~e.~it does
not increase with the dimension of the attractor. In fact, our numerical
simulations indicate only a weak dependence of the actual critical coupling 
strength on the delay time. Last but not least our approach clearly
demonstrates that the success of the synchronisation is independent of
the number of positive Lyapunov--exponents, even if our coupling uses one
scalar variable only, illustrating the results of Stojanovski et al. 
\cite{Stojanovski97}. 

We acknowledge discussions with U.~Parlitz, K.~Pyragas, and Th.~Meyer.

\end{document}